\documentclass{article}
\usepackage{epsfig}
\begin{document}
\title{\bf Scalar curvature of systems with fractal distribution functions}
\author{Marcelo R. Ubriaco\thanks{Electronic address:ubriaco@ltp.uprrp.edu}}
\date{Laboratory of Theoretical Physics\\Department of Physics\\University of Puerto Rico\\R\'{\i}o Piedras Campus\\
San Juan\\PR 00931, USA}
\maketitle
\begin{abstract}
Starting with the relative entropy for two close statistical states we  define the metric  and calculate the scalar curvature $R$ for systems with  classical, boson and fermion fractal distribution functions with moment order parameter $q$. In particular, we find that for $q\neq 1$  the scalar curvature is closer to zero implying that the fractal bosonic and fermionic systems are more stable than the standard ones. 
\end{abstract}
\vspace{0.2in}
Keywords: Scalar Curvature, fractal systems, stability\\
PACS:05.20.-y, 02.40.-k, 02.50.-r, 05.45.Df

\section{Introduction}
From the theory of fractals \cite{F} we learned that given a statistical weight $\Omega(q,\delta)$  of a system with
order parameter $q$ and resolution $\delta$, the fractal dimension is defined as the exponent $d=D_q$ which will make the product $lim_{\delta\rightarrow 0}\Omega(q,\delta)\delta^d$ finite.
With use of  the definition of the Boltzmann entropy  $S(q,\delta)=\ln\Omega(q,\delta)$, the relation between the entropy and the fractal dimension $D_q$ is given by
\begin{equation}
D_q=-\lim_{\delta\rightarrow 0}\frac{S(q,\delta)}{\ln\delta}.
\end{equation}
Based on these definitions and with use of the Boltzmann's H theorem, the generalized entropy and distribution functions for
classical and quantum gases were calculated in Ref.\cite{BD}. The average number of particles with energy $\epsilon$ was shown to
be given by
\begin{equation}
<n(\epsilon)>=\frac{1}{[1+\beta(q-1)(\epsilon-\mu)]^{1/(q-1)}+a}\label{n},
\end{equation}
where $a=0$ for the classical case, and the values $a=-1$ and $a=1$ correspond to Bose-Einstein and Fermi-Dirac cases, respectively. For $q=1$, Equation (\ref {n}) becomes the standard textbook result for classical and quantum ideal gases.
The distribution functions in Equation (\ref {n}) were also obtained in Ref.\cite{BDA} by considering a dilute gas approximation to the partition function of a non-extensive statistical mechanics originally proposed in Ref.\cite{T}.
It is our purpose to study some of the geometric properties of systems with average particle number according to Equation (\ref {n}). The idea of using geometry in thermodynamics is not new \cite {Ti}-\cite{AN}, and  several
authors developed formalisms to measure the distance between equilibrium states through the definition of a metric and the
calculation of the corresponding scalar curvature as a measure of the interactions \cite{I}-\cite{R}.
Some of the applications include classical and quantum gases \cite{R1}\cite{NS}\cite{JM1}\cite{BH}, magnetic systems \cite{JM2}-\cite{JJK}, non-extensive statistical thermodynamics \cite{T-B}\cite{PPP}\cite{O}, anyon gas \cite{MH1}\cite{MH2}, fractional statistics \cite{MH3} and deformed boson and fermion systems \cite{MH4}. Some of the basic results of these approaches include the relationships between the metric  with the correlations of the  stochastic variables, and  the scalar curvature $R$ with the stability of the system, and the facts that  the scalar curvature $R$ vanishes for the classical ideal gas,   $R>0 (R<0)$ for a boson (fermion) ideal gas, and it is singular at a critical point.
Here, we wish to study systems with an average particle number given in Equation (\ref {n}). In Section \ref{FM} we briefly describe the formalism of systems with fractal distribution functions  as reported in Ref.  \cite{MRU1}. In Section
\ref{Me} we obtain the metric from the second order term in the expansion of  the relative entropy between two close statistical states,
and in Section \ref{Scalar} we use the metric to compute the scalar curvature for the classical, Bose-Einstein and Fermi-Dirac cases for $q\neq 1$. In Section \ref{C} we summarize our results.

\section{Fractal models} \label{FM}

\subsection{Classical case}
We will use the short notation
\begin{equation}
\rho_l=[1+(q-1)\beta(\epsilon_l-\mu)]^{1/(q-1)}.
\end{equation}
The probability density is defined
\begin{equation}
\rho=\frac{1}{Z_{MB}}\prod_{l=0}\frac{1}{n_l!}\rho_l^{-n_l},
\end{equation}
where the partition function
\begin{eqnarray}
Z_{MB}&=&\prod_{l=0}\sum_{n_l=0}\frac{1}{n_l!}\rho_l^{-n_l}\nonumber\\
&=&\prod_{l=0}e^{\rho_l^{-1}}\label{ZMB}
\end{eqnarray}
From the definition of the average number of particles with energy $\epsilon_l$ 
\begin{equation}
<n_l>=\frac{\sum_{n_l=0}\frac{1}{n_l!}n_l\rho_l^{-n_l}}{\sum_{n_l=0}\frac{1}{n_l!}\rho_l^{-n_l}},
\end{equation}
we find after summing the series that
\begin{equation}
<n_l>=\rho_l^{-1}.
\end{equation}
as required. In the thermodynamic limit we write for the average
total number of particles $<N>$
\begin{equation}
<N>=\frac{4\pi V}{h^3}\left(\frac{2m}{\beta(q-1)}\right)^{3/2}\int_0^{\infty}\frac{x^2 dx}{[1+x^2-(q-1)\beta\mu]^{1/(q-1)}},\label{N}
\end{equation}
leading to the expression
\begin{equation}
<N>=\frac{-2\pi V}{h^3}\left(\frac{2m}{\beta(q-1)}\right)^{3/2}\frac{1}{[1-(q-1)\beta\mu]^{1/(q-1)-(3/2)}}S,
\end{equation}
where $S$ is a series independent of $\beta\mu$, 
given by
\begin{equation}
S=C_0+\sum_{i=1}^{\infty}\frac{(-1)^m}{m!}\left(\frac{1}{2}\right)...\left(\frac{3}{2}-m\right)C_m,
\end{equation}
with $C_m=\frac{1}{-1/(q-1)-m+(3/2)}$. Solving Equation (\ref{N}) we find that the fugacity $z=e^{\beta\mu}$
has a temperature dependence given by
\begin{equation}
\ln z=\frac{1}{q-1}\left\{1-\left[\frac{-2\pi V}{h^3}\left(\frac{2m}{(q-1)\beta}\right)^{3/2}S\right]^{1/\omega}\right\},\label{lnz}
\end{equation}
where $\omega=\frac{5-3q}{2(q-1)}$. From Equation (\ref{lnz}) we see that the fugacity is restricted to the interval 
$0<z<e^{1/(q-1)}$, which serves as a cut-off that avoids a negative average occupation number.
The correct definition of the average energy is given by
\begin{equation}
<\epsilon>=\frac{4\pi V}{h^3} \int_0^{\infty}\frac{p^2}{2m}<n(p)>^qp^2 dp,\label{E}
\end{equation}
leading, with Equation (\ref{N}), to the required classical result $<\epsilon>=\frac{3}{2}<N>kT$ \cite{MRU1}.
We should remark, that Equations (\ref{ZMB}) and (\ref{E}) are related by the standard definition
\begin{equation}
<\epsilon>=-\frac{\partial\ln Z_{MB}}{\partial\beta}.
\end{equation}
\subsection{Boson case}
Similarly to the classical case,  the definition
\begin{equation}
<N>=\sum_{j=0}\frac{\sum_{n_j=0}n_j\rho_j^{-n_j}}{\sum_{n_j=0}\rho_j^{-n_j}},
\end{equation}
leads to the average occupation number
\begin{equation}
<n_j>=\frac{1}{\rho_l-1}
\end{equation}
with  the probability density and the partition function \footnote{We should remark that our partition function differs from that reported in Ref. \cite{BDA}.}
\begin{eqnarray}
\rho&=&\frac{1}{Z_{BE}}\prod_{j=0}\rho_j^{-n_j},\\
Z_{BE}&=&\prod_{j=0}\sum_{n_j=0}\rho_j^{-n_j}\nonumber\\
    &=& \prod_{j=0}\frac{1}{1-\rho_j^{-1}}.
\end{eqnarray}
As in the standard, $q=1$, Bose-Einstein case the chemical potential is negative.
\subsection{Fermion case}
For the Fermi-Dirac case the average occupation number
\begin{equation}
<n_j>=\frac{1}{\rho_l+1}
\end{equation}
is obtained by defining
\begin{eqnarray}
\rho&=&\frac{1}{Z_{FD}}\prod_{j=0}\rho_j^{-n_j},\\
Z_{FD}&=&\prod_{j=0}\sum_{n_j=0}^1\rho_j^{-n_j}\nonumber\\
    &=& \prod_{j=0}(1+\rho_j^{-1}),
\end{eqnarray}
and the requirement that the average occupation number $<n_j>\in[0,1]$ leads to restrict
the fugacity to the interval  $0<z<e^{1/(q-1)}$.
For Bose-Einstein and Fermi-Dirac cases we obtain 
\begin{equation}
-\frac{\partial\ln Z}{\partial\beta}=<\epsilon>+\sum_{l=0}\sum_{k=1}^{\infty}\left(\begin{array}{c}1-q\\k\end{array}\right)<n_l>^{q+k}\epsilon_l(-a)^k.
\end{equation}
\section{The metric} \label{Me}
For the three cases discussed in the previous section
we cannot adopt any of the standard definitions for the metric like for example \cite{JM1}
\begin{equation}
g_{\alpha\gamma}=\frac{\partial^2\ln Z}{\partial\beta^{\alpha}\partial\beta^{\gamma}},\label{g}\;\;\;\;\beta^1=\beta\;;\;\beta^2=-\beta\mu
\end{equation}
which is valid for exponential distributions. \\The relative entropy, $H(p||P)=\sum_ip_i \left(-\ln(\frac{P_i}{p_i})\right)^{\mu}$, is a very useful concept.  For example, for two close distribution functions $p(x)$ and $p(x+\Delta)$ we can obtain a Fisher's information measure 
\begin{equation}
I_{\mu}=\int dx\left(\frac{dp/dx}{p(x)}\right)^\mu\frac{dp}{dx},
\end{equation}
as part of the second order term in $\Delta$  for the entropic form, with $\mu=1$, $S=-\sum_i p_i\ln p_i$ \cite{V}, and $S=\sum_i p_i(-\ln p_i)^{\mu}$ \cite{MRU2} where  $\mu$ is a fractional parameter. By defining
$\phi(x)=p^{\frac{1}{1+\mu}}$ and considering the Fisher information as a lagrangian density leads to linear and nonlinear differential equations for $\mu=1$ and $\mu\neq 1$, respectively.

Here, based on work in Ref.\cite{I}
we expand the relative entropy for $\mu=1$ between two close densities $\rho(\beta)$ and
$\rho(\beta+d\beta)$ up to second order in $d\beta^{\alpha}$.  Therefore, the information distance
$I(\rho(\beta),\rho(\beta+d\beta))$ between the two close states is written
\begin{equation}
I(\rho(\beta^{\alpha}),\rho(\beta^{\alpha}+d\beta^{\alpha}))= Tr\rho\left(\ln\rho(\beta^{\alpha})-\ln \rho(\beta^{\alpha}+d\beta^{\alpha})\right),
\end{equation}
such that expanding the second order term gives for the metric
\begin{equation}
g_{\alpha\gamma}=\frac{\partial^2\ln Z}{\partial\beta^{\alpha}\partial\beta^{\gamma}}+Tr\rho\sum_{l=0}<n_l>\frac{\partial^2\ln\rho_l}{\partial\beta^{\alpha}\partial\beta^{\gamma}}.\label{gq}
\end{equation}
A simple inspection shows that in the $q\rightarrow 1$ limit Equation (\ref{gq}) reduces to Equation (\ref{g}).
Equation (\ref{gq}) can be simplified leading to the three corresponding metrics:
\begin{eqnarray}
g_{\alpha\gamma}^{MB}&=&\sum_{l=0}\frac{1}{\rho_l^3}\frac{\partial\rho_l}{\partial\beta^{\alpha}}\frac{\partial\rho_l}{\partial\beta^{\gamma}},\\
g_{\alpha\gamma}^{BE}&=&\sum_{l=0}\frac{1}{\rho_l(\rho_l-1)^2}\frac{\partial\rho_l}{\partial\beta^{\alpha}}\frac{\partial\rho_l}{\partial\beta^{\gamma}},\label{gb}\\
g_{\alpha\gamma}^{FD}&=&\sum_{l=0}\frac{1}{\rho_l(\rho_l+1)^2}\frac{\partial\rho_l}{\partial\beta^{\alpha}}\frac{\partial\rho_l}{\partial\beta^{\gamma}},\label{gf}\\
\end{eqnarray}
which can be summarized in the general formula 
\begin{equation}
g_{\alpha\gamma}=\sum_{l=0}\frac{<n_l>^2}{\rho_l}\frac{\partial\rho_l}{\partial\beta^{\alpha}}\frac{\partial\rho_l}{\partial\beta^{\gamma}}.
\end{equation}
Writing in general $\rho_l=[1+(q-1)\sum_{\alpha}\beta^{\alpha}F_l^{\alpha}]^{1/(q-1)}$ we find 
\begin{eqnarray}
\frac{\partial}{\partial\beta^{\alpha}}\frac{\partial}{\partial\beta^{\lambda}}\ln Z&=&\sum_{l=0}\rho_l^{2-2q}F_l^{\alpha}
F_l^{\lambda}\left(q<n_l>-a<n_l>^2\right)\\
g_{\alpha\lambda}&=&\sum_{l=0}\rho_l^{2-2q}F_l^{\alpha}
F_l^{\lambda}\left(<n_l>-a<n_l>^2\right)
\end{eqnarray}
\section{Scalar curvature}\label{Scalar}
\subsection{Classical case}
It has been shown that the scalar curvature \cite{R1}\cite{NS} vanishes for the standard case, but it is tempting to speculate whether that is also the case for $q\neq 1$ . In the thermodynamic limit,with $x=\beta\epsilon$,  we write for example
\begin{equation}
g_{11}=\frac{2}{\sqrt{\pi}}V\beta^{-2}\lambda^{-3}\int_0^{\infty}\frac{x^{5/2}dx}{[1+(q-1)(x+\gamma)]^{\frac{2q-1}{q-1}}}\;\;
\end{equation}
where hereafter $\gamma=-\beta\mu$. This integral converges for $\frac{5-3q}{2(q-1)}>0$, restricting the values of $q$ to the interval $q\in[1,5/3)$.
With use of the integral representation of the $\Gamma$-function
\begin{equation}
\Gamma(y)=w^y\int_0^{\infty}t^{y-1}e^{-wt}dt,\;\;y>0\;;\;w>0, \nonumber
\end{equation}
we obtain for the components of the metric tensor
\begin{eqnarray}
g_{11}&=&V\beta^{-2}\lambda^{-3}h_{5/2},\nonumber\\
g_{12}&=&V\beta^{-1}\lambda^{-3}h_{3/2},\\
g_{22}&=&V\lambda^{-3}h_{1/2},\nonumber
\end{eqnarray}
where the function 
\begin{equation}
h_{\lambda}=\frac{2}{\sqrt{\pi}(q-1)^{(\lambda+1)}}\frac{\Gamma(\lambda+1)\Gamma(\frac{q}{q-1}-\lambda)}{\Gamma(\frac{2q-1}{q-1})}\frac{1}
{[1+(q-1)\gamma]^{\frac{q}{q-1}-\lambda}},
\end{equation}
satisfies
\begin{equation}
\frac{\partial h_{\lambda}}{\partial\gamma}=-\lambda h_{\lambda-1}.
\end{equation}
As is well known \cite{Wei}, the scalar curvature  is given by
\begin{equation}
R=\frac{2}{det g}R_{1212},
\end{equation}
where $det g=g_{11}g_{22}-g_{12}g_{12}$ and the non-vanishing part of the curvature tensor $R_{\alpha\beta\gamma\lambda}$ is given in terms of the
Christoffel symbols

\begin{equation}
R_{\alpha\beta\gamma\lambda}=g^{\eta\theta}\left(\Gamma_{\eta\alpha\lambda}\Gamma_{\theta\beta\gamma}-\Gamma_{\eta\alpha\gamma}\Gamma_{\theta\beta\lambda}\right).
\end{equation}
A simple calculation leads to the result
\begin{equation}
R=\frac{V^{-1}\lambda^3}{4 (det g)^2}\left(5h_{1/2}h_{3/2}^2-6h_{1/2}^2h_{5/2}+h_{3/2}h_{-1/2}h_{5/2}\right)\label{Rc},
\end{equation}
such that after replacement of the definition of the function $h_{\lambda}$ we get that the scalar curvature for the classical fractal case is identically equal to zero. Therefore, in this case the parameter $q$ does not play any role as far as correlations are concerned.

\begin{figure}
\begin{center}
\epsfig{file= 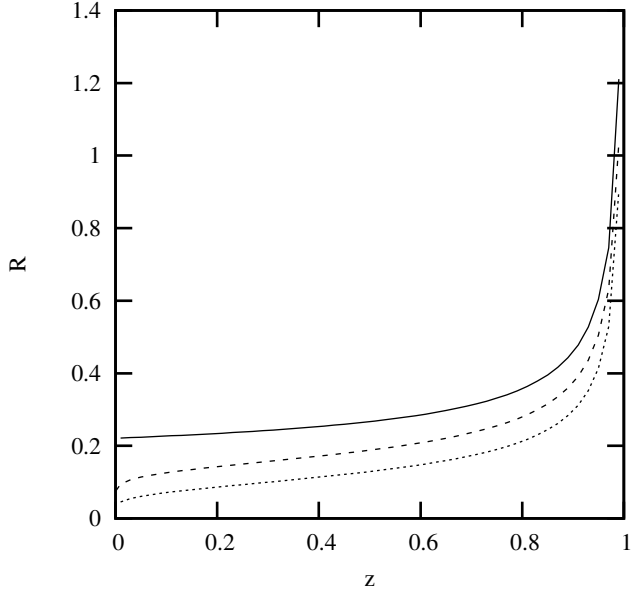,bbllx=50pt,bblly=120pt,bburx=430pt,bbury=250pt}
\end{center}
\caption[]{The scalar curvature $R$, in units of $\lambda^3V^{-1}$, as a function of the fugacity $z$ for bosons at constant $\beta$ for the cases of $q=1$ (solid
line), $q=1.1$ (dashed line) and $q=1.2$ (dotted line).}
\end{figure}

\subsection{Boson and fermion cases}
Here, in order to evaluate the corresponding scalar curvatures we need to replace the summations in Equations
(\ref {gb}) and (\ref {gf}) by integrals
\begin{equation}
G_{\lambda}^{\pm}=\frac{2}{\sqrt\pi}\int_0^{\infty}\frac{x^{\lambda}\Omega^{\frac{3-2q}{q-1}}}{(\Omega^{\frac{1}{q-1}}\pm 1)^2}dx,
\end{equation}
where the $+$ sign is for fermions and the $-$ sign for bosons, and the function $\Omega=1+(q-1)(x+\gamma)$.
In particular, the metric component $g_{11}$ is written
\begin{equation}
g_{11}=V\lambda^{-3}\beta^{-2}G_{5/2}^{\pm},
\end{equation}
and its integral converges for $1\leq q<5/3$.

The functions $G_{\lambda}^{\pm}$ also satisfy
\begin{equation}
\frac{\partial G_{\lambda}^{\pm}}{\partial\gamma}=-\lambda G_{\lambda-1}^{\pm},
\end{equation}
and thus the corresponding equations for $R$ are equivalent to Equation (\ref {Rc}) with the
replacement of the function $h_{\lambda}$ by the functions  $G_{\lambda}^{\pm}$.
 Figures 1 and 2 show  the results of a numerical calculation of the scalar curvature $R$ as a function of the fugacity $z$ for the 
parameter values $q=1,1.1,1.2$ for boson and fermions respectively.
\begin{figure}
\begin{center}
\epsfig{file= 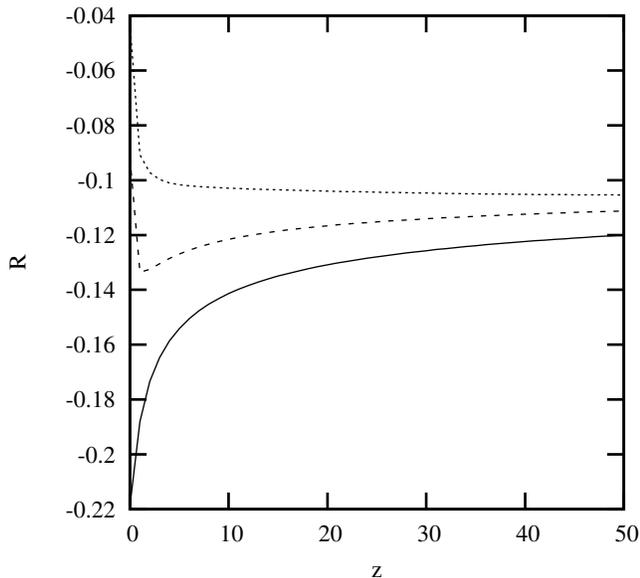,bbllx=50pt,bblly=120pt,bburx=430pt,bbury=250pt}
\end{center}
\caption[]{The scalar curvature $R$, in units of $\lambda^3V^{-1}$, as a function of the fugacity $z$ for fermions at
constant $\beta$ for the cases of $q=1$ (solid
line), $q=1.1$ (dashed line) and $q=1.2$ (dotted line).}
\end{figure}
\section{Conclusions}\label{C}
In this paper, starting from the relative entropy for two close statistical states  we defined the metric for systems
with fractal distribution functions with order parameter $q$. We calculated the scalar curvature $R$
and found that it vanishes for the classical ideal gas, as in the standard case. Numerical calculations for the boson and fermion systems show that the corresponding values of $R$ as a function of the fugacity $z$ are closer to zero than those
in the $q=1$ case, implying that the departure from the value $q=1$ makes the systems more stable. Therefore, for $q\neq 1$
bosons will be less attractive and fermions less repulsive that their standard counterparts. 
Our results are in agreement with those obtained
in a previous work \cite{MRU4} wherein  we showed that  long-range correlations
for the fractal Bose case decrease when the  parameter $q$ departs from the standard value $q=1$.
On the other hand, if one wishes to consider
the order parameter $q$ as a non-extensive parameter it has to be within the context of considering these fractal systems as a dilute approximation to non-extensive statistical mechanics, which consists in replacing the Tsallis partition function by a
factorized one. This type of approximation  has been shown \cite{WL} to be  good  outside a temperature interval that shifts to higher values of $T$ when the number of energy levels increases.  Our results also show  that the sign of $R$ remains unchanged as a function of $z$ implying that these systems do not exhibit anyonic behavior, a fact that  looks impossible to check by performing  an expansion for $z\approx 0$ to obtain  the second virial coefficient
because the partition function is a function of $\ln z$. In addition, our results contrast with the cases of systems with quantum group symmetry where the parameter $q$ interpolates between bosons and fermions in two and three dimensions \cite{MR3}.
\section*{Acknowledgements}
I am grateful to the anonymous reviewers for their comments and constructive criticism to improve the original manuscript.

\end{document}